\documentclass[conference]{IEEEtran}

\IEEEoverridecommandlockouts

\usepackage{cite}
\usepackage{dsfont}

\ifCLASSINFOpdf
\usepackage{graphicx}

\graphicspath{{./Images/}}
\DeclareGraphicsExtensions{.eps, .pdf,.jpeg,.png}
\else
\fi

\usepackage[cmex10]{amsmath}
\usepackage{siunitx}
\usepackage{cite}
\usepackage{psfrag}
\usepackage[utf8]{inputenc}
\usepackage[T1]{fontenc}
\usepackage{amsmath,amsfonts,amsbsy,amssymb}
\usepackage{mathabx}
\usepackage{mathrsfs}
\usepackage[nolist]{acronym}
\usepackage{tabularx}
\usepackage{amssymb}
\usepackage{amsmath}
\usepackage{graphicx}
\usepackage{cite}
\usepackage{multirow}
\usepackage{wasysym}
\usepackage{multirow}
\usepackage{float}
\usepackage{color}
\usepackage{algorithmic}
\usepackage{textcomp}

\begin{document}

\title{Event-based Electricity Metering: An Autonomous Method to Determine Transmission Thresholds
}

\author{\IEEEauthorblockN{Mauricio de Castro Tomé\IEEEauthorrefmark{1}, Pedro. H. J. Nardelli\IEEEauthorrefmark{2}\IEEEauthorrefmark{1}  and Hirley Alves\IEEEauthorrefmark{1} }
	\IEEEauthorblockA{\IEEEauthorrefmark{1}Centre for Wireless Communications (CWC), University of Oulu, Finland\\
	}
	\IEEEauthorblockA{
		\IEEEauthorrefmark{2}Laboratory of Control Engineering and Digital Systems, Lappeenranta University of Technology, Finland\\
	}
	Firstname.lastname@oulu.fi, Firstname.lastname@lut.fi 
}

\maketitle
\thispagestyle{empty}
\pagestyle{empty}

\begin{abstract}

This paper provides an in-depth analysis of the event-based metering strategy proposed by Simonov \textit{et al}. 
This strategy is an alternative to the traditional periodic (time-based) metering where the power demand is averaged in fixed time periods (e.g. every 15 minutes). 
The event-based approach considers two thresholds that trigger an event, one related to the (instantaneous) power demanded, other to the accumulated energy consumed. 
The original work assumed these thresholds fixed for the measurements. 
Our present contribution relaxes this assumption  by proposing a method to set the thresholds from the percentage of the peak power consumption over the period under analysis. 
This approach, in contrast to the time-based and the fixed thresholds, better captures the actual power demanded when different households with diverse power demand profiles are studied. In this sense, our method provides a more efficient way to store electricity demand data while maintaining the estimation error (in relation to the real-time power demand) under acceptable values.
Numerical examples are presented to illustrate the advantage and possible drawbacks of the proposed method.
This paper will appear in the First Workshop on Enabling Energy Internet via Machine type Wireless Communications, IEEE VTC-Spring, 2018.
\end{abstract}

\section{Introduction}
The application of the so-called Internet of Things (IoT) in energy systems has been pointed as a key driver for innovation \cite{DeloitteIoT-energy}, allowing for a more decentralized management structure \cite{bale2015energy}.
If we focus on the electricity power grid, such a modernization processes impose particular challenges, as identified in \cite{nardelli2014models}.
For example, the number of households in central Europe that are becoming capable of generating and storing their locally produced energy is consistently growing.

Virtual power plants, energy communities, virtual utilities, aggregators, micro-grid traders are emerging in this new context, also covered by the EU Winter Package \cite{hancher2017eu}. 
Such new arrangements presuppose a much more active participation in the demand-side; this is generally coined as demand-side management \cite{Palensky2011DemandLoads}.
This requires high quality data for estimation, prediction and planning purposes.
In IoT-based systems, the following steps are usually present: data acquisition, data aggregation/fusion and data analytics (\textit{e.g.} \cite{da2014internet}).

In this paper, we focus on the data acquisition and aggregation steps by assessing two different metering strategies in households, namely time-based and event-based.
As to be further described in the next sections, the first is the traditional periodic metering where the power demanded is averaged for a predetermined time period, i.e. periodic sampling.
The event-based metering was first introduced by Simonov \textit{et al}.\cite{simonov2013event} based on the idea of ``send-on-delta''\cite{Miskowicz2006Send-On-DeltaStrategy} and other asynchronous strategies for wireless sensor networks.
This strategy was further developed by Simonov in a sequence of papers \cite{Simonov2014HybridGrid,Simonov2017GatheringMetering}.
We have also employed the event-based approach to study the effects of communication outages in the signal reconstruction as in \cite{Tome2016JointUsers}, showing that the event-based metering was usually able to obtain the same signal reconstruction quality with less samples (following our previous results \cite{Nardelli2016MaximizingConstraints,alves2016enhanced}, where only time-based metering was analyzed).
Those results, however, have been based on fixed thresholds that trigger the events. 
As expected, these thresholds related to the ``instantaneous'' power demanded and the accumulated energy consumed ultimately determine the performance of this metering strategy.
If two different households have very different consumption patterns, fixed thresholds may be very efficient for one while very poor for the other.
In this case, an automatic method for defining these thresholds in relation to a given sampling period in the time-based metering becomes very important.

In this paper, we propose such a method and test it based on a publicly available electricity consumption database, namely REDD database \cite{kolter2011redd} . 
The database is comprised of six residential customers which were measured during a few days each between April and June of 2011.
The resolution of the power measurements from the house electricity mains is roughly one second, which suits our needs and was the main reason for selecting it. 
Our results show the advantages of using the proposed method by greatly reducing the amount of samples (in comparison to the other strategies) to achieve a consistent good reconstruction (in relation to the high resolution signal) for the houses analyzed here. 

The rest of the paper is divided as follows.
Section \ref{sec_metering} details the two different approaches for data metering.
Section \ref{sec_method} the proposed method is presented and its performance evaluated.
Section \ref{sec_concl} concludes this paper, indicating possible next steps for this research.

\section{Metering strategies}
\label{sec_metering}

\begin{figure}
	\includegraphics[width=1\columnwidth]{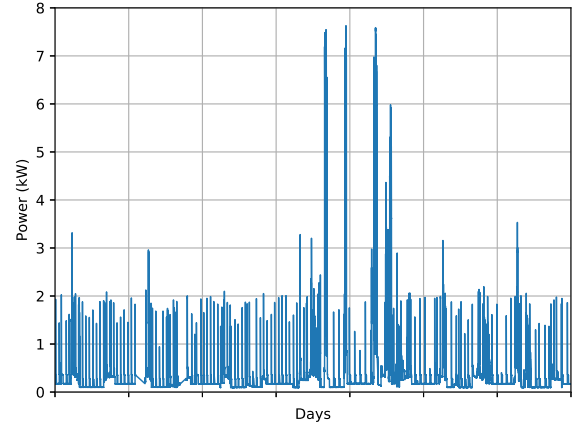}
    \caption{Example of household power demand taken from the REDD database}
    \label{REDDexample}
\end{figure}

Some operational decisions about the power grid management (\textit{e.g.} electricity demand prediction, time-based billing) depend on the daily curves of power consumption in households. 
Figure \ref{REDDexample} illustrates an average power demanded during a typical day by a household with 1-second granularity.
Visually within this time-frame, many spikes can be seen, indicating for example fridge cycles.

Although this information might be useful for some application (\textit{e.g.} abnormalities' detection), most operational decisions by the system operator do not require such small granularity.
In this case, two approaches for metering may be taken: (1) time-based, and (2) event-based.
Both have their own advantages and drawbacks, and heavily depend on the thresholds that define the data sampling, as explained in the following.

\subsection{Time-based metering}

Traditionally, electricity metering is measured at predetermined regular intervals.
This would range from monthly readings (still used in many regions) up to every ten minutes via advanced metering infrastructure (AMI, the sometimes called smart meters).
In other words, metering data related to average power demanded (or energy consumed) is sent every $\Delta t$, where $t$ is a measure of time (seconds, minutes or hours).
The lowest possible $\Delta t$ from the database REDD is one second, which is assumed the basis of comparisons (i.e. "the real-time consumption").
If $\Delta t > 1$s, then the average power demanded is computed as the average power demanded $P$ within the period from $t - \Delta t$ and $t$.

\begin{figure}
	\includegraphics[width=1\columnwidth]{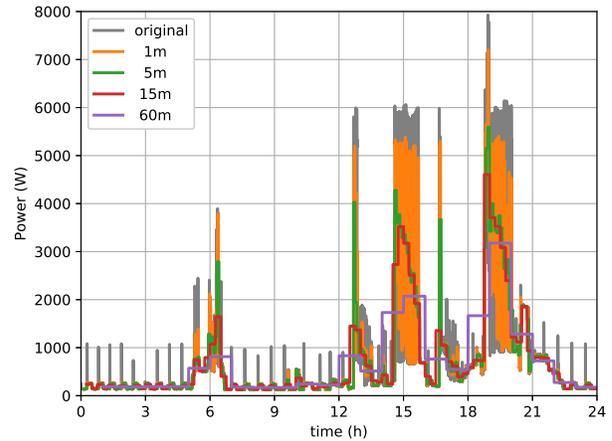}
    \caption{Example of time-based average power reconstruction}
    \label{timebased}
\end{figure}

Figure \ref{timebased} presents a typical daily household profile with $\Delta t = 1,5,15,60$ minutes granularity.
As expected, the larger $\Delta t$, the smoother the curve shape, averaging out possible events like a high power demand.
To cope with this, it is possible to define events that trigger samplings. 
This strategy is known as event-based metering, as discussed next.

\subsection{Event-based metering}

In the event-based metering strategy, as proposed by \cite{simonov2013event,Simonov2014HybridGrid,Simonov2017GatheringMetering}, the samples are not evenly spaced, being taken when some criteria is met, namely:
\begin{enumerate}
\item \textbf{Power variation:} When a given threshold of instant power variation $\Delta P$ is exceeded.
\item \textbf{Energy consumption:} When a given amount of accumulated energy $E$ is reached.
\item \textbf{Time since last measurement was sent:} When a given time $T$ since the last measurement sent has passed.
\end{enumerate}

Extended discussions about these criteria can be found in Simonov's work \cite{simonov2013event}.
Briefly explained, the strategy relies on the fact that during most of the time, the power demanded by an appliance (or for a house, in the case) stays constant. 
So \textbf{1)} uses this fact to capture the (big) transitions in power demand, \textbf{2)} sends periodical information based on the amount of energy consumed (which also helps to track below-threshold power changes), and \textbf{3)} guarantees that even when the power demand (and for consequence, the energy consumption) is low, some periodic information is still sent. It is important noting that, for our analysis, the time component is not considered.

\begin{figure}
	\includegraphics[width=1\columnwidth]{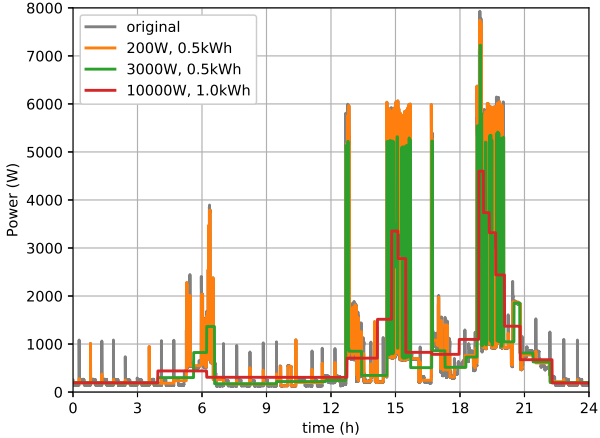}
    \caption{Example of event-based average power reconstruction}
    \label{eventbased}
\end{figure}

Figure \ref{eventbased} shows an example of different signal reconstructions based on this strategy. 
It is possible to see that the first pair of ($\Delta P, E$), in orange color, captures most of the transitions, while the second pair (in green) gets only some major transitions. Finally, the last pair of parameters (in red) is set so high that no event is registered, and only the energy-based measurements are sent.


\section{Proposed method and numerical analysis}
\label{sec_method}

\begin{figure}
	\includegraphics[width=\linewidth]{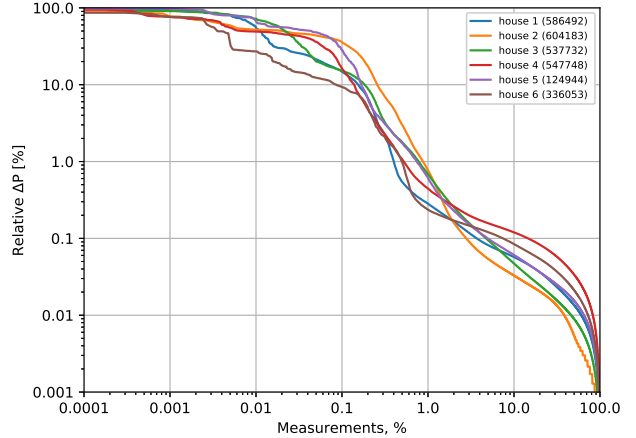}
    \caption{Normalized instant power variation (first differences) from all houses. X-axis: Percentage of measurements; Y-axis: Normalized power variation ($\Delta$P)}
    \label{REDDdiff}
\end{figure}

To set a suitable threshold, we need to first have a better idea on how the instant power variation behaves along the period under analysis.
Figure \ref{REDDdiff} shows the instantaneous relative power differences (\%$\Delta P_\textup{max}$) for the mains power of the six houses (numbered from 1 to 6) from the REDD database, sorted and normalized for the time window used (one week), in \textit{log-log} scale. 
The absolute values for the total energy consumption for the period, the peak power and the peak instant power variation are shown in the Table \ref{info}.
We can see from this plot that the houses' instant power variation behave in a similar fashion (within a certain range), which could be used to give us a lower bound on how many measurements to expect from a given value of the threshold $\Delta P$. 
For the analysis, we decided to not use data from houses 5 and 6 since there are too many missing measurements (House 5 has only about 20\% of its measurements, while House 6 has only about 55\%), which might hinder the analysis.

\begin{table}[b]
\centering
\caption{Information about the houses from the REDD database}
\begin{tabular}{cccc}
House & Peak $P$ [W] & Peak var. [W] & Energy $E$ [kWh] \\
\hline
1 & 7629.07 & 5962.49 & 60109.49 \\
2 & 3253.07 & 2331.04 & 38068.06 \\
3 & 8059.92 & 5640.39 & 61709.76 \\
4 & 4105.19 & 2908.52 & 68196.70 \\
5 & 4901.68 & 3062.39 & 14703.42 \\
6 & 7686.62 & 7328.30 & 49178.99 \\
\end{tabular}
\label{info}

\end{table}

It is possible to see that the majority of the instantaneous power variation for all the houses fall below 1\% of peak instant power variation. 
Conversely, just a few percent of the measures are 1\% or higher, so we decided to set 1\% of the peak power change as the starting threshold for the event-based measurements.
The values chosen for the relative instantaneous power variation were $P = (1, 2, 5, 10, 20, 50, 100) \%$, rounded to $\lceil P_{max} \rceil$, (in KW), where $\lceil x \rceil$ denotes the ceiling ("rounded up") value of x.
The accumulated energy consumption follows a similar rule, using $E = (1, 2, 5, 10, 20, 50, 100) \%$ of rounded to $\lceil E_{mean, daily} \rceil$, (in KWh).
The time-based measurements chosen for comparison were $\Delta t = (10, 30, 60, 300, 600, 900, 1800, 3600, 7200)$ [s].
For each one of the measuring points in both techniques, the accumulated energy within that period is divided by the period, resulting in an average power demand which is compared to the original signal.
As performance metric, we chose to use the Normalized Mean Average Error (NMAE), defined as:
\[\textup{NMAE} = \frac{\sum_{t=1}^{T} |P_{o,t} - P_{m, t}|}{\sum_{t=1}^{T} P_{o,t}},\]
where $P_{o,t}$ is the original power demand for a given time $t$ and $P_{m,t}$ is the corresponding measure from the meter in each one of the cases.
We opted to use NMAE instead of a mean squared error metric (such as RMSE) because the latter puts a heavy penalty in large power deviations that happen in the averaging process of the time-based strategy, while being captured in more detail with the energy-based technique.

Figure \ref{house1_error}, top part, shows the NMAE of the event-based measurements (colored lines with markers) in comparison with the error for the time-based measurements (gray dashed lines). 
Each colored line corresponds to one accumulated energy threshold, whereas the markers correspond to the power variation thresholds. 
On the bottom part, we have the corresponding amount of measurements for each one of the measurement strategies for the period of one week (or 604800 seconds). The dashed gray lines correspond to the reference values for the selected time-based strategies.
This way we can compare with relative ease how each one of the measuring strategies perform in relation to the others.

\subsection{House 1}
\begin{figure}[ht]
\includegraphics[width=\linewidth]{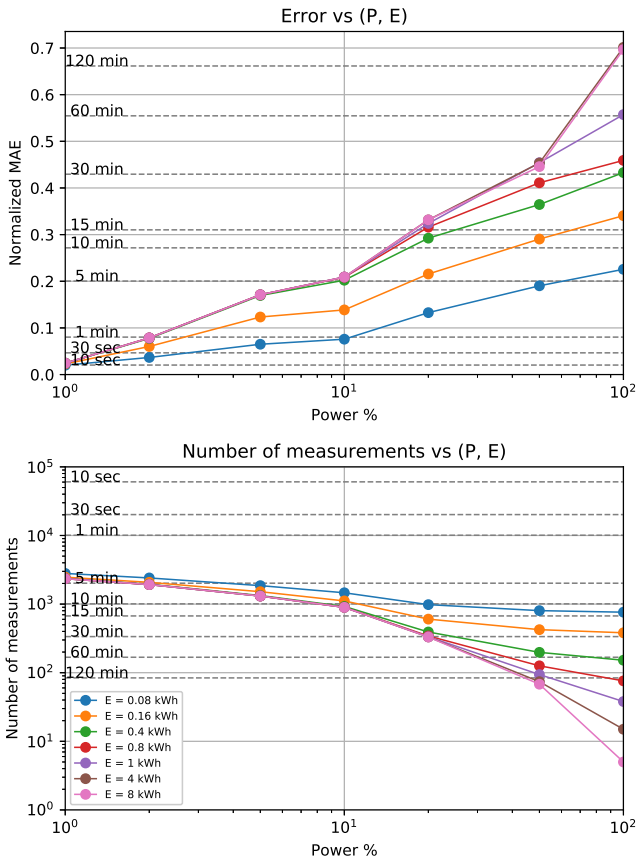}
\caption{House 1. Top: NMAE; Bottom: Number of measurements}
\label{house1_error}
\end{figure}

As an example, consider the points corresponding to $E=1\%$ (blue line), $P=10\%$ (fourth point from the left), for House 1 in Figure \ref{house1_error}.
It is possible to see on the top side of the plot that this point has a NMAE error comparable to a 1-minute sampling rate, while on the bottom side of the plot, it has a daily amount of measures (read: information \textit{transmitted, processed} and \textit{stored}) lower than a 5-minute sampling rate. That roughly translates to send information with similar quality using 80\% less data. It is also worth noting that the lower the power threshold, the better it performs.


\subsection{House 2}
\begin{figure}
\includegraphics[width=\linewidth]{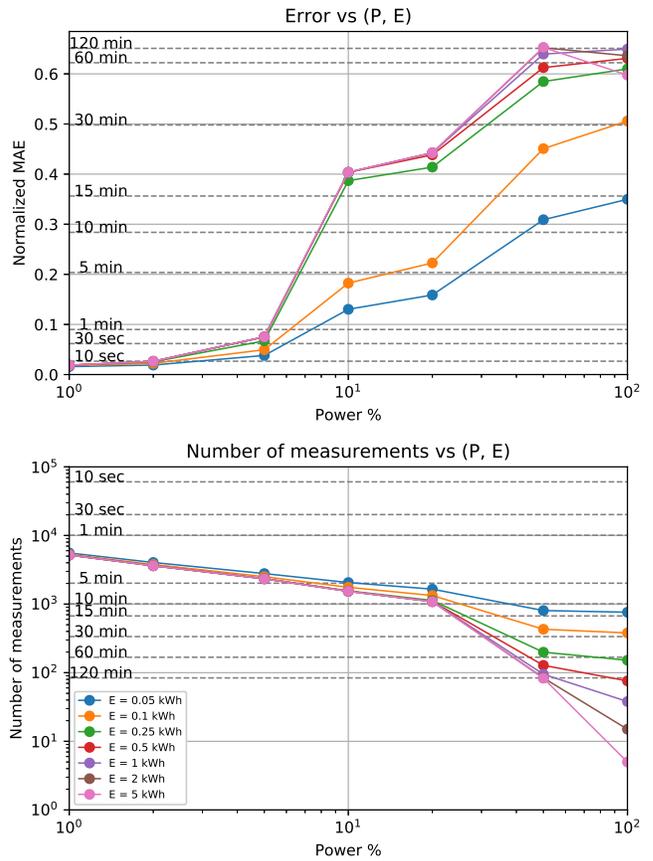}
\caption{House 2. top: NMAE; bottom: Number of measurements}
\label{house2_error}
\end{figure}

In Figure \ref{house2_error}, House 2 shows similar behavior for lower power thresholds, but the performance of the event-based technique degrades quickly, in particular when we raise the energy limit. This is probably due to the low level of activity within the house, besides being the lowest consumption of all houses. 
%

\subsection{House 3}
\begin{figure}
\includegraphics[width=\linewidth]{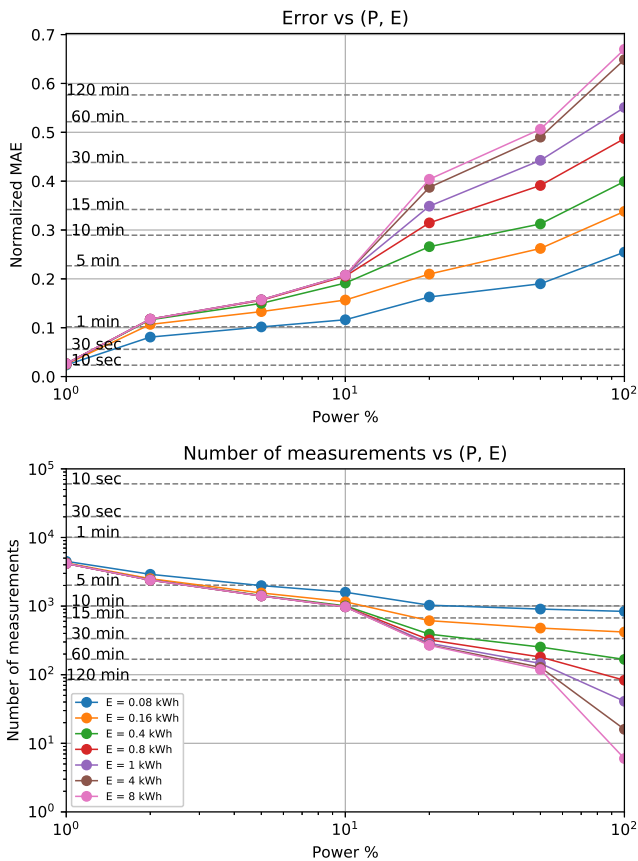}
\caption{House 3. top: NMAE; bottom: Number of measurements}
\label{house3_error}
\end{figure}

House 3, as can be seen in Figure \ref{house3_error}, behaves in a similar fashion to House 1, although the energy thresholds seem to have more influence on the error on the latter case.

\subsection{House 4}
\begin{figure}
\includegraphics[width=\linewidth]{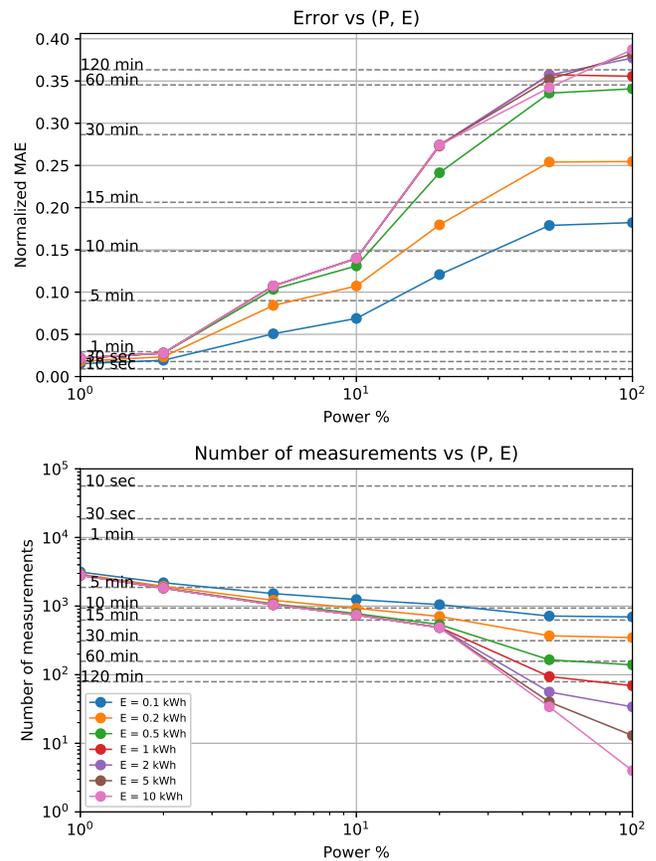}
\caption{House 4. top: NMAE; bottom: Number of measurements}
\label{house4_error}
\end{figure}

House 4 behaves in a slightly different way than the other ones, in the sense that it starts off relatively worse than all of the previous ones in terms of comparative performance with the time-based measurements, as we see in Figure \ref{house4_error}. This is probably due to the fact that this house has a long period in which the base consumption is kept almost constant, which influence an be seen on the error metric. While most of the houses have a worst-case NMAE at around 0.65, House 4 has its worst-case NMAE around 0.35.
In essence, this means that its event-based performance, although slightly worse than the comparative cases in the time-based strategy, is still acceptable.

\subsection{All Houses}
From the bottom part of the plots from all the houses it is possible to see that even though the event-based strategy is not deterministic (in the sense that it is not possible to determinate the amount of measurements taken beforehand), using the relative values for the thresholds of power and energy makes all the houses have their measures roughly  within the same range. 
This somewhat consistent behavior is especially true when we are using values in the lower range of $\Delta P$ or $E$ (about 10\% or less).

The compression characteristics of the event-based metering is especially interesting, since it means less energy for transmission and less measurements to store. 
Using the lowest threshold for $\Delta P$ in our simulations (1\%), compression ratios of 13:1 to 26:1 were attained in comparison with the 10s average power, with similar NMAE. 
If a squared error metric was used to penalize for big deviations, the results would be even more favorable towards the event-based approach.

\section{Conclusions and future work}
\label{sec_concl}

The proposed method was shown to be useful to determine consumption-dependent thresholds for the values of instant power variation and energy consumption for the event-based strategy, consistently outperforming the time-based strategy. 
Following the same procedure for the different houses, we were able to greatly reduce the amount of measures needed to reconstruct the signal maintaining a very good resolution, even when compared to a very high time-based resolution. 
We expect to improve the proposed method and test it using other databases, including also industries and commercial consumers, as well as data from different countries.
Our plan is to implement the strategy in real environments, where the thresholds shall be determined via analysis of historical data and users' own inputs (\textit{e.g.} holidays).

\section*{Acknowledgements} 
This work is partially supported by Aka Project SAFE (Grant n.303532) and Strategic Research Council/Aka BCDC Energy (Grant n. 292854).

\bibliographystyle{IEEEtran}
\bibliography{ref}

\end{document}